# WOMEN'S SOLIDARITY AND SOCIAL MEDIA: SISTERHOOD CONCEPT IN #LASRESPONDONAS, A FACEBOOK GROUP IN PERU

Juan Bossio, Communications Department, Pontificia Universidad Católica del Perú, jfbossio@pucp.pe

Illari Diez, Communications Department, Pontificia Universidad Católica del Perú, illari.diez@pucp.pe

**Abstract:** Women in Peru are exposed daily to gender violence and exclusion. Several feminist groups have used social media to share information, debate, denounce, organize, and provide help to victims. This contribution analyzes the meaning of female solidarity, sisterhood or *sororidad*, as a feminist political concept among members of one feminist Facebook group. We reviewed the how various feminist and sisterhood concepts clustered together in the group communications by reviewing group publications and comments as far as interviewing key former and current members. The sisterhood concept was found to be central to feminist practice.

## 1. INTRODUCTION

This paper is about the meaning of female solidarity, or sisterhood, and how it is related to other feminist concepts among members of a feminist Facebook group. Analytical constructs employed are the evolution of Latin American and Peruvian feminism, characteristics of virtual communities, and the use and appropriation of social media by social movements.

Firstly, though we refer to a feminist movement, we recognize there are many forms of feminism, with their corresponding ways to think, live, or act. In Latin America, feminist movements reflect specific cultural, ethnic, social, and linguistic diversity with particular characteristics that encompass demands beyond the fight for gender equality (Rivera, 2018; Sardiña, 2020). This diversification began in the 20th century, but in recent decades has passed through important changes. In the 1990s, there predominated a liberal feminism that gained some gender equality rights for women, but was adapted to neoliberalism (Coba & Herrera, 2013). At the beginning of the 21st century, these feminist movements began to address globalization and included a critique of American Westernization, along with racism and the aftermath of colonialism (Gargallo, 2007).

Thus, Peruvian feminist movements have passed through three waves. From 1911 to 1933, feminist women were focused on the fight for the right to vote, women's access to equal educational opportunity, and support for other causes like the rights of workers and indigenous people (Barrientos & Muñoz, 2014). From 1973 to the present, the feminist struggle focused on a demand for equal citizenship in every aspect. During the 1970s, women held street demonstrations to gain visibility (Barrientos & Muñoz, 2014) and institutions arose to address women's workplace issues (Vargas, 1985). Starting around 1978, new autonomous groups appeared and joined the left to find solutions to women's problems from a Marxist perspective (Vargas, 1985). In the 1980s, women embraced a more general sense of rights. During the 1990s feminist movements became complex and fragmented, bringing new forms of expression (Vargas, 2004). The third wave, arising in parallel to the second wave, from 2003 to the present, saw a definition of feminism in terms of sexuality. Although, the Peruvian state promoted many norms that support equal opportunities, such as protections against domestic violence and sexual assault, this didn't result in practical outcomes in key aspects of women's lives, including domestic and sexual violence, and abortion (Barrientos & Muñoz, 2014). In this context, groups that defended sexual diversity rights adopted themes of feminist radicalism. There is an ongoing debate about whether to defend gender identities when fundamental conditions of women haven't reached an acceptable minimum (Barrientos & Muñoz, 2014). Between the second and third waves there are coinciding themes regarding legal abortion, violence, patriarchy, child pornography, etc. Questions that divide feminists have to do with the relation (or not) between patriarchy and capitalism, the inclusion of trans women as feminist subjects, regulation or prohibition of sex work, and other issues.

Within these feminisms, one concept present and established as a practice is the notion of sisterhood. The word used by feminists in Spanish —where sisterhood and brotherhood share the same term: *hermandad*— is *sororidad*. *Sororidad* could thus be understood as the construction of relationships of complicity, mutual support, and solidarity among women to generate support networks across broad areas of life (Torcuato et al., 2017). Further, from Lagarde's perspective, *sororidad* is defined as ethical, political, and practical aspects of contemporary feminism. It urges women to seek positive relationships and political alliances with other women, with an aim to conduct specific actions against oppression and provide support to gain empowerment (Lagarde, s.d.). *Sororidad* implies that women who practice it must see each other as equals (Lagarde, 1990, 2001 as cited in Ojeda, 2013).

The case presented in this paper is a Facebook group called #LasRespondonas (#LR) created in 2016 by feminist women in Lima, Peru. It gets its name from statements of the former cardinal Cipriani calling women ministers who promoted free distribution of emergency contraceptive pills "Respondonas" or "Backtalkers" (Diez, 2020a). It was created as a space for feminist debate and as a project incubator, moreover, #LR was considered a *sororidad* space above all else. This centrality is the reason why this case was chosen among others such as *Bolsa de Trabajo Feminista - Peru, aló compita?* or *Feministas buscando casa - Peru*. The group grew rapidly to 20,000 members in a few months. Some anticipated projects didn't materialize, and others happened separately. Since 2018, the group has been used for feminist debate, accusations, and requests for help (Diez, 2020a).

Secondly, online communities have been conceptualized as associations of people who interact around shared interests in a non-private way, using a common language and led by shared norms (Agostini & Mechant, 2019; Brown et al., 2007; Jin et al., 2010). The shared interests could be shaped around social problems, as well as other common issues (Suazo et al., 2011). These communities can have different size or scope (Preece, 2001). Preece considers an online community to be "any virtual social space where people come together to get and give information or support, to learn or to find company" (p. 348). These virtual spaces could be of many kinds, such as online chat rooms, social networking sites, and weblogs, among others (Ridings et al., 2002). Organizations, including online communities, emerge or "drift" rather than following preestablished plans, as the environment changes (Ciborra et al., 2000). Consequently, the value of the community isn't from the platform itself, it is contained in the information, ideas, and content shared by members during discussions and interaction (Jin et al., 2010). Another important aspect of online communities is the 'online homophily', understood as the shared demographic characteristics of its members, such as gender, age, and education (Ruef et al., 2003). The appearance of agents with extreme beliefs who primarily interact among themselves is more likely to happen in large networks or communities (Madsen et al., 2018). In other words, it leads to the formation of clans that bring comfort to members and avoid disagreements (Törnberg, 2018).

Thirdly, from a socio-technical perspective, artifacts are socially constructed, situated, and contingent (Siri, 2008). Systems and platforms are designed for some purpose, but users reshape them by processes of adaptation (Walsham, 2001). The social change agent is not the technology itself, but the uses and sense construction built around it (Grint and Woolgar, 1997). Participatory culture preexists Social Network Systems (SNS), but they have been enhanced and reshaped by how SNS are used (Jenkins, 2006). Since the 1970s, online communities have connected people who share interests, feelings, or ideas (Rheingold, 2000); they have rapidly adapted and transformed over the interval (Brown et al., 2007). One example of their capacity for adaptation is the rise of social movements on these platforms.

Social movements are groups of people that face collective challenges to common objectives; there is solidarity among their members and an interaction with authorities, opponents and elites (Tarrow, 1994). An important part of this interaction is through public representations of worthiness, unity, numbers, and commitment (WUNC) (Tilly & Woods, 2009). Requests for recognition, including that of feminists, has been a feature of demands from social movements, and are even more important nowadays (Castells, 2012; Melluci, 1996; Tilly & Woods, 2009; Treré, 2015). Today's movements make an intensive use of Internet resources and platforms, especially social media, thus becoming net movements (Castells, 2012; Jenkins et al., 2016; Postill, 2018). Indeed, it should be noted that feminism is one of the first international movements to appropriate IT (Gajjala y Oh, 2012). Social movements act through repertoires (Tilly & Woods, 2009); some have evolved or adapted to digital terrains as "digital *escrache*" (denunciation campaigns on social media) or slogans rendered in hashtags. Others are new, such as video mapping, hashtag campaigns like #MeToo, or hashtag crashings like those recently made by Kpoppers (Bossio, 2020; Jenkins et al., 2016; Treré, 2015). Furthermore, digital media is increasingly used to share opinions, coordinate actions, learn tactics for street action (e.g., tutorials on how to defuse tear gas canisters), or to express or give solidarity (Bossio, 2020; Treré, 2015).

The relevance of this article is supported by two principal aspects. The first one is the context of the case study. Peruvian women are daily exposed to gender violence and unequal access to work, education, etc. For example, in the first two months of this year, there have been 23 femicides and 51 femicide attempts (MIMP, 2021). The second one is the scarcity of studies of the use of digital media by feminist movements and *sororidad* in Latin America in general and Peru in particular. For example, in Peru, Caballero (2018) showed how a group of Facebook was used to connect offline networks, which facilitate the adherence of people and the emergence of the 'Ni una menos' mobilization. Furthermore, Soto (2019) described how in that Facebook group women could share sensitive disclosure and generate *sororidad*. Finally, Diez (2020a, 2020b) studied the transition of *sororidad* from online to offline spaces through feminists interaction in #LasRespondonas and the security construction to guarantee it. The absence of major studies on these convergent topics highlights the need to contribute visibility to positive and negative aspects of the use of social media by feminist movements. Then, this paper seeks to provide an exploratory approach -from an specific case- to the use of digital media for discusing *sororidad* and other feminist political concepts in Latin America.

## 2. METHODOLOGY

The main question of this article concerns to how is *sororidad* understood and how this concept relates to others among members of a feminist Facebook group. Using a qualitative netnographic approach (Hine, 2015; Pink et al., 2015), we combined interviews of key participants and analysis of Facebook posts and commentaries.

The following categories were defined to organize the research and answer questions about:

- *Sororidad* definition by activists,
- Subjects and beneficiaries of *sororidad*,
- The place of *sororidad* among other concepts,
- Ways in which *sororidad* happens,
- Absence of *sororidad* in feminist interaction,
- *Sororidad* and social media,
- *Sororidad* as a political tool.

## 2.1. Data and sample selection

The key respondents to the interviews were chosen for their knowledge of the #LasRespondonas Facebook group (#LR). The first informant was Nani Pease, a founder and former member of the group; the second was Laura Balbuena, another founder; and the third was Mariana Velasco, a moderator. Is important to highlight that they allowed to be named.

Posts to analyze were identified by searching on key terms related to *sororidad* and feminism found in the Facebook group. The combinations of ideas used were: 'political *sororidad*', 'feminism and *sororidad*', 'political feminism', '*sororidad*', 'is not *sororidad*', 'victim and *sororidad*' and '#YoTeCreo' (believe the victim).

Criteria for choosing posts to analyze included date of publication (posts from 2018 onwards), relevance or relation to the defined categories, and discussion content. Excluded were posts with few relevant comments, reposts, or publicity.

Honoring the commitments to moderators and the group, we sought authors permissions to include their posts and comments in the sample. All authors were contacted via messages through the platform. One author communicated that she didn't want her posts and comments to be considered, but the majority of authors whose content wasn't considered did not reply. From the 16 publications and 380 comments initially selected from 98 women, we were authorized to use 143 posts and/or comments.

## 2.2. Materials

The interview guide asked about respondents' experiences at #LR, related *sororidad* and feminist concepts, how *sororidad* is carried out in practice, its role in political action, actors and roles in *sororidad* formation, and how it is shaped by social media.

The analysis of posts and comments was carried out using a list of concepts related to *sororidad*, with the aim to identify which of them appear connected to it during the group debates, either explicitly or implicitly. The list of concepts was created considering the known theory and an assessment of the content. Related concepts arose during the analysis. The final list of categories included: patriarchy/sexism, deconstruction, empowerment, intersectionality, ethics, representation, activism, #YoTeCreo (hashtag in support of the claims of a female victim of violence), sex work, normalization, commodification/objectification, stigmatization, respect, empathy, security/privacy, confidence, acknowledgment, sisterhood (*hermandad*), help, and feminist debate (how to debate within the group).

The first level of analysis allowed us to identify the most common concepts and to exclude 14 comments that lacked a connection to *sororidad*. The analysis presented in the next section considers 129 items (12 posts and 117 comments). Finally, four posts and their comments were selected for a content analysis because of the richness of the debate they contain including all the most considered concepts. This analysis focused on the definition of *sororidad* encountered, how the concepts identified were used, and limits to *sororidad*.

## 3. RESULTS

For context, this section begins with an introduction to the Facebook group and its purposes. Then it looks at the concepts related to or aligned with *sororidad*, followed by how the concept is understood at #LasRespondonas. Those results are based on how *sororidad* is a political concept and tool for feminism. Finally, we consider the relation between social media use and *sororidad* practice.

## 3.1. Evolution, functionality and debate on #LasRespondonas

According to interviewees, the group adapted and evolved, taking on a life of its own. As it grew, the founders asked members to become moderators. #LR was originally a group mostly populated by cisgender women, but because of the presence of transgender women and some debates, trans moderators were included. Another change was to the type of group; initially it was an open group, but when some people started taking screenshots or sharing sensitive information (testimonies of violence, identity of victims, etc.) it was decided to make it private. As well, moderators campaigned to eliminate suspicious profiles and enforce the admission process. To gain access, women had to be added by a group member and answer screening questions.

As stated above, #LasRespondonas is a space of *sororidad*, which becomes evident when one of the group members posts a request for help. Some of the situations that generate requests are gender violence, depression, unwanted pregnancy, and economic problems (Diez, 2020a). When a request is posted, group members begin to offer helpful comments. For example, they make supportive statements, add to the visibility of the post using the word 'up' so more people can see it, narrate personal experiences to show emotional support, give needed information or share where it can be found, give advice, and offer direct help (Diez, 2020a). While in some circumstances help provided on the post is enough to solve a problem, in others tangible support like clothes, money, pills for pregnancy interruption, accommodation, etc. are needed. In that case, group members usually communicate through a private space, like Facebook Messenger or WhatsApp to coordinate a way to meet in person (Diez, 2020a).

This interaction generates relationships between the women of the group. They begin to know each other personally, identify commonalities, build friendships, empathize, and give more help. This is possible because there is a feeling of confidence: they feel that they will always receive help when needed. Also, they feel that they won't be judged when talking about their problems.

Nevertheless, at times the debate becomes intense, and it would appear there is no *sororidad*. Although as Laura said, "conflict is needed in order to grow", friction arises when some kind of consensus, like the right to abortion, is questioned. While those shared ideas are not negotiable, moderators believe that abrasive reactions work against *sororidad* because they silence participants who may leave without having the chance to change their views. They believe there should be empathy and understanding toward those who come to a feminist space, and the context of their ideas should be considered. As Nani, who left #LR after dozens of such debates, said: "there should be the right to not be clear, not have a manual with all the answers, or to share doubts". Other tough debates arise when views from different sides of the feminist movement collide over issues like domestic or sex work. When those debates intensify, *ad hominem* remarks pop up—for instance, "procurer" for those who advocate regulations for sex work.

Such attacks and interventions about how the debate among feminists should proceed were coded on our sample as "Feminist Debate". This became the most common category, which partially describes around 30% of posts or comments analyzed. On the other hand, combating violence against women was the issue that produced unity out of diversity.

### 3.2. Concepts related to *sororidad*

The interviewees identify several concepts as being related to *sororidad*: feminism, solidarity, empathy, respect, fellowship, mutual care, avoid judging the victim, not revictimize, recognition of the other, recognition of differences, deconstruction, confidence, believe the victim, and sisterhood.

Apart from "Feminist Debate", the most common themes were respect, empathy, intersectionality, patriarchy/sexism, believe the victim (#YoTeCreo) and sisterhood; these were followed by deconstruction, commodification/objectivization, sex work, stigmatization, help, acknowledgment, activism, ethics, and normalization. Some themes would be associated with feminist theory as patriarchy, a system ruled by men; intersectionality, the consideration of social facets that combine to produce discrimination; deconstruction, the process of critiquing a person's own beliefs/feelings/thinking associated with living in a patriarchal society; commodification, valuing women as things; stigmatization or normalization, seeing violence or other women´s problems as normal. The rest concern political or practical issues, some related to collective action, such as believing the victim and activism, others to the building of relationships among feminists or women as respect, sisterhood, acknowledgment and ethics, or to active *sororidad* empathy and help.

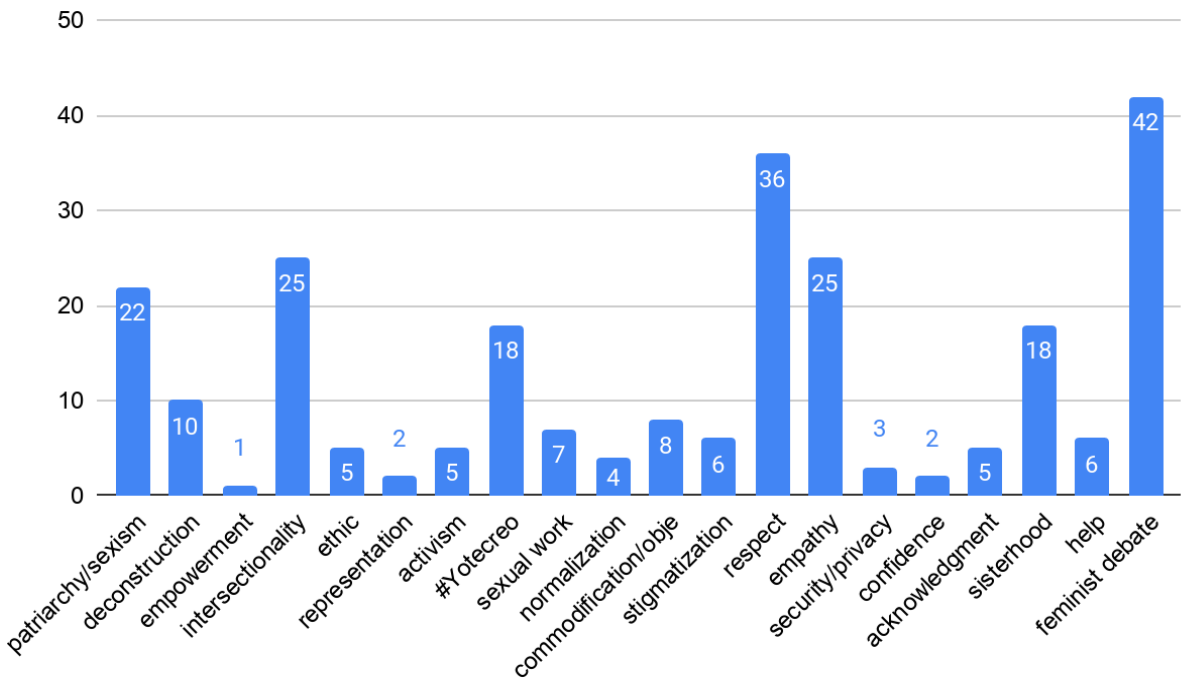

**Figure 1. Categories assigned to post and comments**

The concepts above are related to *sororidad* in different ways in feminist debates. For instance, one post looking for a live-in maid opened a debate about whether this work regime was enslavement of women and whether seeking it could be considered anti-feminist. Some said that those who hire women under this regime are using them to be able to personally grow themselves; others argued that this regime was helpful for women who live far from their workplace, etc. Also, there were two forms of 'debate': a respectful one, asking for empathy, and another that used aggressive language and sarcasm. This second form of debate was indicated as an example of lack of *sororidad*. In a similar way, another post debated about sex work and *sororidad*; it was said that exploiting or humiliating women is not *sororidad*. So, it can be observed that *sororidad* is linked to intersectionality, empathy, and the attitude and actions on a debate.

Another example was provided by a post which used the Medusa myth to discuss topics such as patriarchy, commodification, and *sororidad*. There are two sides of the myth. On one hand, Athena represents a patriarchal woman, who doesn't practice *sororidad* because she judged and punished Medusa for being a sexual assault victim. Moreover, Medusa is perceived as a monster when she is the victim. On the other hand, Athena is seen as a wise person, who made Medusa ugly so that men would leave her alone. This perspective highlights that in a patriarchal society where it has been normalized that beauty can attract harassment (commodification of women) and female victims of sexual assault are stigmatized, being ugly becomes a protection. Then, *sororidad* is linked to support and non-judgment of women victims of violence.

### 3.3. How is *sororidad* conceived in #LR?

Solidarity among women is commonly referred to in English as sisterhood. But since it is a core feminist concept in Spanish as *sororidad*, we choose to use that term. It implies actions such as help, support and care, and feelings of sisterhood and fellowship. As a product of feminist history and action, *sororidad* can be understood from a political and/or theoretical perspective. The construct of *sororidad* allows members to introspect to understand and act as allies for outward-facing actions. While *sororidad* is conceived as a feminist practice, it encompasses all women.

To perform *sororidad* requires actors to empathize, and as a consequence avoid behaviors like judgment, exposure, or evaluation of a victim. *Sororidad* is to tell an unknown woman: "you are not alone", to show care for the other, and create a shared space of confidence. *Sororidad* implies seeing other women around you as allies and not as competitors.

However, there can be tensions within a group when *sororidad* is not a generally shared value. All interviewees declared that *sororidad* should be universal (it shouldn't be conditioned by the qualities or history of those who need it), but they also realize that it is a goal. As mentioned above, some tense debates drift far from the spirit of *sororidad*, but there are also deliberate arguments against universal *sororidad*. Moderators realize that some #LR members think that they can't act with *sororidad* towards right-wing women, religious believers, patrons of housekeepers, policewomen, xenophobes, women that didn't act with *sororidad* in the past, intolerant women or transgender women. There is a mixture of feelings and ideologies supporting these exclusions. From one perspective, this could be explained as a misunderstanding of feminism; from another, there are context issues or life experiences that give rise to this. Both moderators suggest an empathic approach to these exclusionist feelings, offering different ways to deal with them. However, there is a controversy between the idea of 'educating' or having a dialogue with those women to see if they can change their minds and excluding them from the group to avoid conflict.

### 3.4. Sorority as a political tool

As mentioned, *sororidad* can also be viewed in political terms. According to Laura, "*sororidad* is feminism in action": without *sororidad* feminism is just a theory, not a political movement. For moderators, the bases of feminism are community feeling, sisterhood, and willingness to help. *Sororidad* is what unites women, what makes them believe in themselves and others. Thus, *sororidad* is viewed as a practice, a feeling, and a central feminist concept.

It helps women to advocate for change as a way of life, and of political collective action that maintains unity in spite of differences. Moreover, *sororidad* is a survival tool for women in a patriarchal society that excludes them, pays them less for their work, doesn't protect them from violence or abuse, and cuts their social benefits—in short, a society that doesn't guarantee the fulfilment of their basic rights. Examples are a lack of publicly funded childcare or access to birth control in a country where abortion is illegal.

*Sororidad* also becomes a political position when women are victims of violence. As Nani points out, the role of women in that context is to take care of the victims, support them, and give them strength during the process of bringing legal charges; in other words, *sororidad* means not to replicate patriarchal violence, revictimize, or judge. To pose questions and investigate the facts is the job of a judge, not of a feminist.

### 3.5. Sorority and digital media

The relationship between social media and *sororidad* depends on who, how, and for what reason media are being used. Social media can become a hell when it provides tools to revictimize women who have suffered violence or generate disputes among allies. On the other hand, it can be a space for discussion and allyship, providing an opportunity for various forms of help: legal, medical, economic, or psychological, in terms of allowing accusations of abuse to spread.

Nani observes that there is an "automatic empathy activation" but affirms that it must be nurtured to become practical. She warns that interactions on social media can produce misunderstandings and needless disputes because of unconsidered reactions and assumptions about the characteristics and motivations of other participants. While responding to content in groups like #LR, members forget they can't really see the reactions of others. Thus, an element of restraint and empathy, trying to think from the other person's perspective, is essential to building a healthy community.

Other important issues concern security and privacy. Internet content is easy to spread but hard to contain. In these cases, it is urgent to protect the identity and testimony of victims. For *sororidad* to manifest on #LR a secure space (understood as "a virtual environment that, while not completely risk-free, is intended to minimize or control risks, and to generate a feeling of security and confidence in members to maintain their well-being when they interact in the space") is required (Diez, 2020b, p.14). Group members then feel they will receive the help they need without being judged. This allows them to ask for help, share information, offer help, and meet face-to-face (Diez, 2020b).

## 4. DISCUSSION AND CONCLUSIONS

This case of study allows a practical understanding that *sororidad* shows the same characteristics in digital spaces and offline spaces. The notion that *sororidad* is conceived around the actions of help, support, care, and the feeling of sisterhood and fellowship can be seen as network construction between women (Torcuato et al., 2017). Similarly, it can be seen as a more political, ethical, and practical form of feminism (Lagarde, s.d.). This is because it is not thought of as simply a glue for feminists and the base of the movement but is understood as feminism itself in action, in the Facebook group and beyond. It leads women to act against patriarchy and promote actions to guarantee the fulfillment of their rights (Lagarde, s.d.). Thus, *sororidad* is a survival tool for women in a system that doesn't fully support them (Lomnitz, 1978).

However, the question of who should be included in the practice of *sororidad* is a delicate one. As discussions have shown, sometimes *sororidad* is lost, or its universality is questioned. This arises from the diversity of feminist movement demands (Sardiña, 2020) and the evolution of the movement itself (Barrientos & Muñoz, 2014), showing that offline political beliefs are reflected in digital spaces.

Accordingly, the #LR group has drifted, coalescing from the multiple goals it had in the beginning, through its explosive growth to its settled/established usages: feminist deliberation, denunciations of perpetrators, and assistance (Ciborra et al., 2000). Furthermore, as its members have accommodated/adapted certain functionalities of the platform to give visibility to urgent announcements by adding visibility functions to certain comments ("up-leveling") so they appear more often on members' walls, they provide an interesting case to prove how people appropriate this tools to its owns interests as said by Walsham (2001). One issue important to moderators and members has been to build a secure digital space for all. Just a space free of revictimization and without filtered information would stimulate women to share their worst experiences and look for help from others (Diez, 2020b). So, moderators changed the group to a secret one and enforced an admission process to avoid the intrusion of aggressors or infiltrators to the extent possible.

Moreover, #LR provides a good case to understand the characteristics of a virtual community as a) it is a large virtual community (Preece, 2001) built around its members' social interests (Suazo et al., 2011; Preece, 2001); b) the public interaction between women in it is led by norms such as to respect each other, not to share information, etc. (Agostini & Mechant, 2019; Brown et al., 2007; Jin et al., 2010); c) its members seem to have an 'online homophily' because the majority of them are women (Ruef et al., 2003), however, they have differences in education or feminist sophistication level, as cited by our interviewees; d) it manifests the construction of a cluster with some extreme beliefs —which is not always open to disagreement— as some members just want to share the space with other people who think as they do (Madsen et al., 2018; Törnberg, 2018). This poses a challenge to moderators: to maintain free and critical digital interaction that is respectful, and to practice *sororidad* among members.

Although, #LR itself is not a social movement, it gives the opportunity to interpret it inside the feminist movement in Peru. Members share challenges and objectives and there is a sense of solidarity (here, *sororidad*) among them (Tarrow, 1994). However, the group hadn't developed towards an organization that produced WUNC demonstrations (Tilly & Woods, 2009). In that way,

#LR is more like a sharing space inside the movement. It could be argued that performing *sororidad* as a political action is a repertoire (Tilly & Woods, 2009).

Finally, #LR case allow to continue proving the increase in use of the digital platform as a political terrain for diffusion of ideas, debate, and organized action, both face-to-face and digital, resulting from widespread use, adaptation, and appropriation of SNS (Bossio, 2020; Castells, 2012; Jenkins et al., 2016; Postill, 2018). Also, it reinforces that identity or 'identities' of intersectionality and gender issues are considered crucial to this kind of groups, as has been reported worldwide (Castells, 2012; Jenkins et al., 2016; Melluci, 1996; Postill, 2018; Tilly & Woods, 2009; Treré, 2015).

In conclusion, since patriarchy can be claimed as a central concept for feminist theory, from the analysis of #LR, it can be argued that *sororidad* is correspondingly central to feminist practice in daily life as in political action.

### 4.1. Limitations and future research

This research has been limited to some extent by concerns for sources and subjects' willingness to collaborate. The sample of posts and comments includes only those who have agreed to be included. That permission wasn't granted for four posts and more than a hundred comments they generated. Thus, a selection bias can be assumed.

Considering the relevance of the discussion on trans inclusion it would be interesting to interview the trans woman moderator of the group. Furthermore, after realizing the importance of how debate is conducted in this virtual space, it is clear that interviewing some people with outlying positions or behavior would have been useful.

Finally, this research opens other questions. How closely tied to Peruvian culture is the way in which these feminists debate? What can be done to promote energetic but not harmful debate? How is *sororidad* conceived in other feminist groups? What else can be said about the debate on the universality of *sororidad*?

## REFERENCES AND CITATIONS